\documentclass[%
 reprint,
 amsmath,amssymb,
 aps,
]{revtex4-2}
\usepackage{multirow}
\usepackage{graphicx}
\usepackage{dcolumn}
\usepackage{bm}
\usepackage{amsmath,amssymb,amsfonts}
\usepackage{mathrsfs}
\usepackage{bbm}
\usepackage{slashed}
\usepackage{verbatim}
\usepackage[T1]{fontenc}
\usepackage[utf8]{inputenc}
\usepackage{hyperref}
\usepackage{ifthen}
\usepackage{forloop}
\usepackage[english]{babel}
\usepackage{mathbbol}
\usepackage[usenames,dvipsnames]{xcolor}
\usepackage[normalem]{ulem}
\usepackage{tabularx, booktabs}
\usepackage{subcaption}

\newcommand{\bea}{\begin{eqnarray}}
\newcommand{\eea}{\end{eqnarray}}
\newcommand{\be}{\begin{equation}}
\newcommand{\ee}{\end{equation}}

\definecolor{darkgreen}{rgb}{0,0.4,0}

\graphicspath{{figures/}}

\begin{document}


\title{Evidence of Taste-Symmetry restoration for K\"ahler-Dirac fermions on Euclidean Dynamical Triangulations}

\author{Mingwei Dai}
\affiliation{Department of Physics, University of Illinois Urbana-Champaign, Urbana, Illinois, 61801, USA}
\author{Walter Freeman}
\affiliation{Department of Physics, Syracuse University, Syracuse, NY 13244, USA}
\author{Mason Grieb}
\email{mpgrieb@syr.edu}
\affiliation{Department of Physics, Syracuse University, Syracuse, NY 13244, USA}
\author{Jack Laiho}
\email{jwlaiho@syr.edu}
\affiliation{Department of Physics, Syracuse University, Syracuse, NY 13244, USA}
\author{Marc Schiffer}
\affiliation{Institute for Mathematics, Astrophysics and Particle Physics, Radboud University, Heyendaalseweg 135, 6525 AJ Nijmegen, The Netherlands}
\author{Judah Unmuth-Yockey}
\affiliation{Infleqtion, 125 S. Wacker Dr., Chicago, IL 60606, USA}

\begin{abstract}
We study the spectrum of two-particle bound states of K\"ahler-Dirac fermions coupled via gravity in Euclidean Dynamical Triangulations (EDT).  We work in the quenched approximation, where the geometry is allowed to fluctuate, but there is no back-reaction from the matter on the geometry.  The K\"ahler-Dirac action respects an exact $U(1)$ symmetry that is related to continuum chiral symmetry, and this $U(1)$ symmetry ensures that some of the bound-state masses vanish in the chiral limit of vanishing fermion bare mass.  The bound-state masses that are not protected by the $U(1)$ symmetry do not vanish in the chiral limit, but the scale associated with the symmetry breaking does not appear to be a physical scale like the Planck mass or the curvature scale of the emergent de Sitter background.  Rather, the mass scale associated with this non-degeneracy appears to be a positive power of the lattice spacing, decreasing as the continuum limit is approached.  This suggests that K\"ahler-Dirac fermions in EDT can be interpreted as the gravity analog of staggered fermions in flat-space lattice field theories, with similar behavior.
\end{abstract}
\maketitle

\section{\label{sec:intro}Introduction}

It is an interesting and open question as to whether lattice field theory approaches may be used to formulate quantum gravity.  Such approaches typically invoke Weinberg's asymptotic safety scenario, which would be realized if there exists a non-trivial fixed point, with a finite-dimensional ultraviolet critical surface of trajectories attracted to the fixed point at short distances \cite{Weinberg:1980gg}.  Since the couplings of the theory are not necessarily small at the fixed point, nonperturbative methods are required, and the lattice provides results with controlled systematic errors in nonperturbative investigations of field theory for problems that are amenable to Monte Carlo methods.  It is also interesting to investigate the less ambitious question of whether one can formulate a lattice theory that correctly reproduces the low energy effective theory of gravity, since the effective theory makes universal predictions that any candidate theory of gravity should reproduce \cite{Donoghue:1994dn, Dunbar:1994bn, Donoghue:1999qh, Bjerrum-Bohr:2002gqz}.  Understanding what is needed for the latter is likely a prerequisite for investigations of the former, more ambitious question.

The main focus of this work is the lattice approach known as Euclidean dynamical triangulations \cite{Agishtein:1991cv, Ambjorn:1991pq}.  Simulations of this formulation find the emergence of de Sitter-like geometries in the large volume limit at fine lattice spacings \cite{Laiho:2016nlp, Asaduzzaman:2022kxz, Dai:2023tud, Dai:2024vjc}, given a particular tuning of the bare parameters and the inclusion of a local measure term.  The Planck scale can be determined reliably from the semiclassical fluctuations about de Sitter, and simulations at a succession of ever finer lattice spacings suggest that there is no barrier to taking a continuum limit \cite{Bassler:2021pzt, Dai:2024vjc}.  The existence of a continuum limit would then establish the existence of an ultraviolet fixed point on the lattice, so this evidence may be taken as cause for optimism for the asymptotic safety scenario.  Additional evidence for asymptotic safety comes primarily from the functional renormalization group (FRG) \cite{Wetterich:1992yh, Reuter:1996cp}, see \cite{Percacci:2007sz, Reuter:2019byg} for textbooks, and \cite{Bonanno:2020bil, Dupuis:2020fhh, Knorr:2022dsx, Eichhorn:2022gku, Morris:2022btf, Wetterich:2022ncl, Martini:2022sll, Saueressig:2023irs, Platania:2023srt, Bonanno2023, Reichert:2020mja, Basile:2024oms, Eichhorn:2026uqj} for reviews and lecture notes, and from other lattice studies \cite{Ambjorn:2005db, Ambjorn:2005qt, Ambjorn:2007jv, Ambjorn:2024qoe}.  We are motivated by these results to extend EDT calculations to include the matter sector, since this is a feature of our world and must be included in a more realistic theory.  Also, there are many studies of matter coupled to gravity using the FRG, see, e.g., \cite{Eichhorn:2012va, Dona:2013qba,  Meibohm:2015twa, Pastor-Gutierrez:2022nki}, and \cite{Eichhorn:2018yfc, Eichhorn:2022gku, Eichhorn:2026uqj} for reviews, so there may be useful points of comparison for lattice calculations.  Exploratory studies within EDT have looked at including scalar fields \cite{Ambjorn:1993ta, deBakker:1996qf, Dai:2021fqb} and quenched K\"ahler-Dirac fermions \cite{Catterall:2018dns}.  The results of the most recent works studying two-particle bound states were consistent with expectations for gravitational interactions \footnote{See also Ref.~\cite{Maas:2025rug} for evidence of a state with mass around one tenth the Planck mass in the pure gravity implementation of causal dynamical triangulations.}.  In Ref.~\cite{Catterall:2018dns}, evidence was presented that K\"ahler-Dirac fermions formulated on EDT geometries recover four degenerate flavors of Dirac fermions in the large volume, flat space limit.  In this work, we revisit this result with new ensembles at finer lattice spacings, many more volumes, and more reliable determinations of the lattice spacings.   

The similarity of earlier results for gravitational bound states of K\"ahler-Dirac fermions \cite{Catterall:2018dns} to those of staggered fermions in flat-space lattice gauge theories makes it tempting to regard K\"ahler-Dirac fermions living on EDT geometries as the lattice gravity analog of staggered fermions.  K\"ahler-Dirac fermions on EDT retain an exact $U(1)$ symmetry that is a subset of the full chiral symmetry, just as in the case of lattice QCD using staggered fermions (in flat space) \cite{Catterall:2018dns}.  In the lattice QCD case there is strong numerical evidence that the flavor symmetry, also known as ``taste'' symmetry when the extra flavors are unwanted, is restored in the continuum limit.  The residual lattice symmetries of the theory also ensure that no fine-tuning is required to restore it.  In unquenched (flat-space) simulations with staggered fermions in four dimensions, a fractional power of the fermion determinant is required in order to perform simulations at something other than an integer multiple of four degenerate-mass fermions.  Though the rooting procedure is non-local, a large body of evidence \cite{Sharpe:2006re, Kronfeld:2007ek, Golterman:2008gt, FlavourLatticeAveragingGroupFLAG:2024oxs}, which includes theoretical arguments and agreement between results from different types of fermion discretizations, suggests that the continuum theory of rooted staggered fermions is in the same universality class as other lattice fermions that do not need to root, such as Wilson fermions.  Might this apply to K\"ahler-Dirac fermions on EDT?  For reason to be skeptical, we point out that the K\"ahler-Dirac fermion action on a curved, classical background is not in the same universality class as (four copies of) Dirac fermions, although they become equivalent in the flat space limit \cite{Banks:1982iq, Catterall:2018lkj}.  The different fermions become nondegenerate in the presence of curvature, since the different flavors couple differently to gravity \cite{Banks:1982iq}.  This is not the whole story, though, because EDT does not assume a classical background; rather the long-distance behavior emerges from a nonperturbative evaluation of the path integral sum over geometries.  The recovery of long-distance physics is highly nontrivial, so it is not {\it a priori} obvious what will happen when K\"ahler-Dirac fermions are coupled to EDT.  In this work we present additional evidence that the two-particle bound states of different K\"ahler-Dirac fermion types become degenerate in the continuum limit, further strengthening the analogy to staggered fermions where taste symmetry is restored.

This paper is organized as follows.  In Section~\ref{sec:edt} we review the EDT formulation, the K\"ahler-Dirac action, and the lattice simulations that were carried out in \cite{Dai:2024vjc}.  In Section~\ref{sec:tasteBreaking} we present our numerical analysis of bound states of K\"ahler-Dirac fermions.  We study the finite-size effects and continuum extrapolation of the various bound state mass-splittings in the chiral limit.  We conclude in Section~\ref{sec:conclude}.

\section{\label{sec:edt}Lattice Formulation}

\subsection{Euclidean Dynamical Triangulations}

The Euclidean quantum gravity partition function is formally given by a path integral over all geometries
\bea\label{eq:part}  Z_E = \int {\cal D}[g]{\cal D}[\overline{\omega}]{\cal D}[\omega] e^{-S_{EH}[g]-S_{KD}[g,\overline{\omega},\omega]},
\eea
where the Euclidean Einstein-Hilbert action is
\bea  \label{eq:ERcont} S_{EH} =  -\frac{1}{16 \pi G}\int d^4x \sqrt{g} (R - 2\Lambda),
\eea
with $R$ the Ricci curvature scalar, $g$ the determinant of the metric tensor, $G$ Newton's constant, and $\Lambda$ the cosmological constant.  We choose for the matter sector the K\"ahler-Dirac action,
\bea \label{eq:KDcont} S_{KD} = \int d^4x \sqrt{g} \ \overline{\omega}(d-\delta + m_0)\omega,
\eea
where the K\"ahler-Dirac fields, denoted $\omega$, consist of a collection of Grassmann valued $p$-form fields that are coupled through the exterior derivative $d$ and its adjoint $\delta$, with $m_0$ the mass \cite{Kahler:1962, Banks:1982iq, Catterall:2018dns}.

In this work we neglect the contribution of $S_{KD}$ in Eq.~(\ref{eq:part}) to the Boltzmann weight when generating the ensembles. This is an uncontrolled approximation commonly referred to as the quenched approximation, and it allows us to use the existing ensembles that were generated previously \cite{Dai:2024vjc}. Even in quenched calculations it is possible to study the properties of valence fermions and their gravitational interactions, with important qualitative features expected to be the same as in the full theory.  

  The discretized version of the Einstein-Hilbert action is the Einstein-Regge action \cite{Regge:1961px}
  \begin{equation} \label{eq:GeneralEinstein-ReggeAction}
S_{ER}=-\kappa\sum_{j=1}^{N_2} V_{2}\delta_j+\lambda\sum_{j=1}^{N_4} V_{4},
\end{equation}
  \noindent where $\delta_j=2\pi-{\cal O}(t_j)\arccos(1/4)$ is the deficit angle around a triangular hinge $t_j$, with ${\cal O}(t_j)$ the number of four-simplices meeting at the hinge, $\kappa=\left(8\pi G \right)^{-1}$, $\lambda=\kappa\Lambda$, and the volume of a $d$-simplex is 
\begin{equation} \label{eq:SimplexVolume}
V_{d}=\frac{\sqrt{d+1}}{d!\sqrt{2^{d}}}a_{\rm lat}^d,
\end{equation}
\noindent where the equilateral $d$-simplex has a side of length $a_{\rm lat}$.  After performing the sums in Eq.~(\ref{eq:GeneralEinstein-ReggeAction}) one finds
\begin{equation}\label{eq:DiscAction}
S_{ER}=-\frac{\sqrt{3}}{2}\pi\kappa N_{2}+N_{4}\left(\kappa\frac{5\sqrt{3}}{2}\mbox{arccos}\frac{1}{4}+\frac{\sqrt{5}}{96}\lambda\right).
\end{equation}
We rewrite the Einstein-Regge action in the simple form 
\bea\label{eq:ER}  S_{ER}=-\kappa_2 N_2+\kappa_4N_4,
\eea
introducing the parameters $\kappa_4$ and $\kappa_2$ in place of $\kappa$ and $\lambda$ for convenience in the numerical simulations.

The path integral of dynamical triangulations is formulated directly as a sum over geometries, without the need for gauge fixing or the introduction of a metric.  The dynamical triangulations approach is based on the conjecture that the path integral for Euclidean gravity is given by the partition function \cite{Ambjorn:1991pq, Bilke:1998vj}
\bea\label{eq:ZE} Z_E = \sum_T \frac{1}{C_T}\left[\prod_{j=1}^{N_2}{\cal O}(t_j)^\beta\right]e^{-S_{ER}},
\eea
where $N_i$ is the number of simplices of dimension $i$, and $C_T$ is a symmetry factor that divides out the number of equivalent ways of labeling the vertices in the triangulation $T$.  The term in brackets in Eq.~(\ref{eq:ZE}) is a nonuniform measure term \cite{Bruegmann:1992jk, Bilke:1998vj}, where the product is over all triangles, and ${\cal O}(t_j)$ is the order of triangle $j$, i.e. the number of four-simplices to which the triangle belongs.  This corresponds in the continuum to a nonuniform weighting of the measure in Eq.~(\ref{eq:part}) by $\prod_x \sqrt{g}^{\beta}$, where we leave $\beta$ an adjustable parameter in the simulations.  It was shown in \cite{Laiho:2016nlp} that $\beta$ must be tuned in order to recover geometries with semiclassical properties in four dimensions.  

\subsection{Details of the simulations}


We briefly review the properties of the lattice geometries and the methods for generating the ensembles used in this work; for more details see \cite{Dai:2023tud, Dai:2024vjc}.  The configurations consist of geometries that are constructed by gluing together four-simplices along their ($4-1$)-dimensional faces.  The four-simplices are equilateral, with fixed edge length $a_{\rm lat}$.  The set of all four-geometries is approximated by gluing together four-simplices, and the dynamics is encoded in the connectivity of the simplices.  We work with a set of degenerate triangulations in which the combinatorial manifold constraints are relaxed.  The set of degenerate triangulations is larger than that of combinatorial triangulations, and the degeneracy allows distinct four-simplices to share the same 5 distinct vertex labels \cite{Bilke:1998bn}.  It was shown that the finite size effects of degenerate triangulations are smaller than those of combinatorial triangulations \cite{Bilke:1998bn}, giving it some advantages over combinatorial triangulations, especially when studying phase transitions of the theory.  A comparison between the two discretizations shows no essential difference in the phase diagram between degenerate and combinatorial triangulations in four dimensions \cite{Ambjorn:2013eha, Coumbe:2014nea}.  The use of degenerate triangulations introduces a complication when formulating K\"ahler-Dirac fermions on the lattice geometries, since the formulation requires that the geometries be orientable. As shown in \cite{Catterall:2018dns}, the class of degenerate triangulations retains this property and the expected relations for the lattice analogues of the operators $d$ and $\delta$ are still satisfied.

The simulations use Markov Chain Monte Carlo methods to evaluate the partition function in Eq.~(\ref{eq:ZE}), where the rejection free algorithm introduced in Ref.~\cite{Dai:2023tud} was used.  This algorithm mimics the Metropolis-Hastings algorithm in that the partition function is sampled using a Markov chain constructed from a series of local moves, but the moves are always accepted, with a weight that is used to reconstruct the proper distribution.  The ensembles generated in \cite{Dai:2024vjc} have been pushed to finer lattice spacings than were previously accessible.  The new algorithm enabled this, since the region of the phase diagram that corresponds to finer lattice spacings is characterized by a very low Metropolis acceptance rate.  The rejection free algorithm led to a speedup of one to two orders of magnitude compared to standard Metropolis in the relevant part of the phase diagram.

As is typical with dynamical triangulations calculations, the simulations were done at approximately fixed lattice
four-volumes. The local moves that evolve the lattices in Monte Carlo time are the same Pachner moves \cite{Gross:1991je} that are used in the standard Metropolis-Hastings algorithm.  The Pachner moves are topology preserving, making it straightforward to restrict the sum over geometries to those with fixed topology, $S_4$ in this case.  The Pachner moves require the volume to vary in order for them to be ergodic, though in practice
the volume fluctuations about the target four-volume are constrained to be small. This is done by introducing a
volume preserving term in the action $\delta \lambda|N_4^f - N_4|$, which keeps the four-volume close to a fiducial value $N_4^f$.  In principle one should take the limit in which $\delta\lambda$ goes to zero, but setting it too small leads to large volume fluctuations and long autocorrelation times.  For the ensembles used in this work, 
$\delta\lambda=0.04$ was chosen.  Once the volume is specified, the parameter $\kappa_4$ is constrained to a particular value. The other parameters, $\kappa_2$ and $\beta$, then form a two-dimensional parameter space in which to explore the phase diagram.  

\begin{figure}
    \centering
    \includegraphics[width=\linewidth]{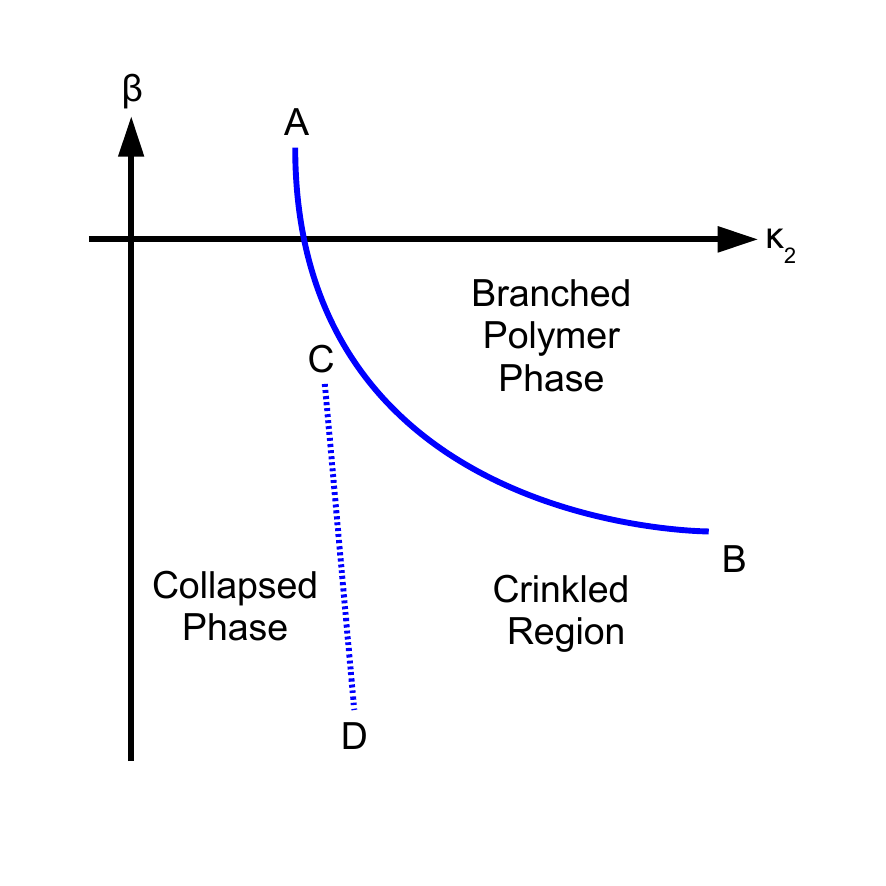}
    \caption{Schematic of the phase diagram as a function of $\kappa_2$ and $\beta$.}
    \label{fig:phaseDiagram}
\end{figure}

The phase diagram has been studied
in previous work \cite{Coumbe:2014nea, Laiho:2016nlp, Dai:2023tud}, and a schematic is shown in Fig. \ref{fig:phaseDiagram}.  There are two unphysical phases, the collapsed phase and the branched polymer phase, which were identified already in early EDT simulations \cite{Agishtein:1991cv, Ambjorn:1991pq, Catterall:1994pg, deBakker:1994zf, Ambjorn:1995dj, Egawa:1996fu}.  The
solid line AB is a first order line that separates
the two phases \cite{Bialas:1996wu, deBakker:1996zx, Ambjorn:2013eha, Coumbe:2014nea, Rindlisbacher:2015ewa}.  The phases are not smooth, with emergent fractal dimensions that differ from four.  One useful definition of fractal dimension is the Hausdorff dimension, which is a measure of geometry and is defined by the scaling of the volume of a sphere with its radius.
The branched polymer phase has Hausdorff dimension 2,
while the collapsed phase has a large dimension that may diverge in the infinite volume limit. The so-called ``crinkled region" is closer to the first-order phase transition line where physical properties begin to emerge.

If one tunes the bare parameters so that the simulations are sufficiently close to the first-order phase transition, approaching it from the left, the resulting geometries resemble four-dimensional de Sitter space \cite{Laiho:2016nlp, Dai:2023tud, Dai:2024vjc}.
Furthermore, if the tuning to the phase transition line is followed out to larger $\kappa_2$, then the lattice spacing gets finer and discretization effects become smaller \cite{Bassler:2021pzt, Dai:2024vjc}.  There appears to be no barrier, other than the cost of the simulations, to going to arbitrarily fine lattice spacings.


\subsection{Lattice ensembles}
\label{sec:ensembles}

In this work we study fermion bound states on a number of tuned (quenched) lattice ensembles across a range of volumes and many different lattice spacings.  Table \ref{tab:ensembles} gives the parameters for the ensembles used here, which are a subset of the ensembles generated for \cite{Dai:2024vjc}.  In some cases we do not use the full configuration set reported in \cite{Dai:2024vjc} for some of our larger lattice sizes, mostly due to the expense.  Improvements to our analysis code could make this more feasible.

Since we are particularly interested in the behavior of bound-state masses in the continuum limit, it is important that we have reliable lattice spacing determinations.  Because our lattices are not hypercubic, there is no single definition of the lattice spacing that is natural for all quantities.  Also, because the geometry at long distance scales is emergent, this leads to one of the most convenient units, the renormalized dual lattice spacing $\ell$, not being simply related via local geometry to the direct lattice spacing $a_{\rm lat}$.  Some quantities are more naturally measured in terms of the link length $a_{\rm lat}$ or in terms of $\ell$, requiring that we know their relative sizes at each nominal lattice spacing (i.e., at fixed $\kappa_2$).  Thus, we need to determine the conversion factor $a_{\rm lat}/\ell$ from the simulations.  We also require the relative lattice spacing $\ell_{\rm rel}$ so that we can convert results at different lattice spacings to a common set of physical units.

The ratio $a_{\rm lat}/\ell$ can be obtained assuming that the emergent geometries match the classical de Sitter space solution, as presented in \cite{Dai:2024vjc}.  The absolute and relative lattice spacings are obtained by considering the semiclassical fluctuations about de Sitter space, and identifying the partition function of the theory with the exponential of the de Sitter entropy, which is the semiclassical result.  This allows us to determine the Planck scale in lattice units, thus allowing us to convert our masses in units of the inverse lattice spacing to units of the Planck mass.  Comparing the Planck mass in lattice units across lattice spacings allows us to determine the relative lattice spacings, which are especially important for the scaling study presented here.  These determinations represent a significant improvement over the earlier effort to study the scaling of fermion discretization effects on EDT lattices in \cite{Catterall:2018dns}.  We take the values of these spacings from \cite{Dai:2024vjc}, where their determinations are described in greater detail.

Table~\ref{tab:ensembles} shows the bare parameters $\kappa_2$ and $\beta$ used in the simulations, as well as the target number of four-simplices, $N_4$.  Once $N_4$ is chosen, $\kappa_4$ is tuned so that the four-volume fluctuates about the target volume.  Table~\ref{tab:ensembles} also gives the number of configurations analyzed for this work on each ensemble. The ratio of the direct to (renormalized) dual lattice spacings $a_{\rm lat}/\ell$ and the relative (renormalized) dual lattice spacings $\ell_{\rm rel}$ are also given in Table~\ref{tab:ensembles}, including the errors in their determinations, for ease of reference.  

\begin{center}
\begin{table}
\caption{Ensembles used in this work.  The first column is the $\kappa_2$ value of a given lattice spacing.  The second is the tuned $\beta$ value of the ensemble.  The third is the target volume in four-simplices of the ensemble. The fourth column is the number of configurations analyzed, and the sixth column is the ratio of the direct to renormalized dual lattice spacing at the given $\kappa_2$ value.  The seventh is the (renormalized dual) relative lattice spacing at the given $\kappa_2$ value. }
\label{tab:ensembles}
\begin{tabular}{c c c c c c} 
\hline \hline
$\kappa_2$& $\beta$ & $N_4$ & $N_{\rm config}$ &  $a_{\rm lat}/\ell$ & $\ell_{\rm rel}$\\
\hline
\multirow{6}{4em} {\centering 2.45}& -0.574 & 8000 & 3023 & \multirow{6}{4em} {\centering 9.02(19)} &  \multirow{6}{4em} {\centering 1.45(12)}\\
& -0.555 & 12000 & 1431  \\
& -0.544 & 16000 & 2160  \\
& -0.530 & 24000 & 1598  \\
& -0.520 & 32000 & 822  \\
\hline
\multirow{6}{4em} {\centering 3.0}& -0.800 & 8000 & 2196  & \multirow{6}{4em} {\centering 11.66(16)} &  \multirow{6}{4em} {\centering 1.00}\\
& -0.782 & 12000 & 3696  &\\
& -0.771 & 16000 & 1719 &\\
& -0.756 & 24000 & 1578  &\\
& -0.746 & 32000 & 1465  &\\
& -0.735 & 48000 & 4722 &\\
\hline
\multirow{9}{4em} {\centering 3.4} & -0.870 & 12000 & 2573 & \multirow{6}{4em} {\centering 16.34(19)} & \multirow{6}{4em} {\centering 0.618(49)} \\
& -0.853 & 16000 & 3857 &\\
& -0.839 & 24000 & 2582  &\\
& -0.830 & 32000 & 2040  &\\
& -0.822 & 48000 & 103  &\\
& -0.815 & 64000 & 125 & \\
\hline
\multirow{9}{4em} {\centering 3.8} & -0.907 & 16000 & 3615  & \multirow{6}{4em} {\centering 23.67(52)} & \multirow{6}{4em} {\centering 0.386(51)} \\
& -0.893 & 24000 & 2299  \\
& -0.880 & 32000 & 2358 \\
& -0.864 & 48000 & 715  \\
& -0.864 & 64000 & 74  \\
& -0.857 & 96000 & 91 \\

\hline\hline
\end{tabular}
\end{table}
\end{center}

\subsection{K\"ahler-Dirac fermions}

The K\"ahler-Dirac equation arises by taking the square
root of the Laplacian operator written in terms of exterior derivatives.  The K\"ahler-Dirac operator is
\bea  K = d - \delta,
\eea
where $d$ is the exterior derivative and $\delta$ its adjoint.  Using the fact that $d d=0$ and $\delta\delta=0$, it follows straightforwardly that
\bea  KK^\dagger = -(d - \delta)^2 = (d\delta + \delta d) = -\square,
\eea
where $\square$ is the Euclidean Laplacian.  This suggests an alternative to the usual Dirac equation,
\bea \label{eq:KD}  (d - \delta + m_0)\omega = 0,
\eea
where $\omega$ is a collection of differential forms.  Since the K\"ahler-Dirac operator and the usual Dirac operator are both in some sense the square root of the scalar Laplacian, we may expect their respective equations to be related.  We can find this relation by considering the matrix formed from a combination of gamma matrices,
\bea \label{eq:psi}  \Psi = A + A_\mu\gamma^\mu + \frac{1}{2}A_{\mu\nu}\gamma^\mu \gamma^\nu + ...
\eea
The action of $\gamma^\mu \partial_\mu$ on the matrix $\Psi$ is identical to the operation of the K\"ahler-Dirac matrix $K$ upon $\omega$ \cite{Banks:1982iq}.  Thus $\Psi$ obeys
\bea \label{eq:DiracPsi} (\gamma^\mu \partial_\mu + m_0)\Psi = 0.
\eea
We can interpret each column of the $4\times 4$ matrix $\Psi$ as a Dirac spinor, so that Eq.~(\ref{eq:DiracPsi}) corresponds to four copies of the Dirac equation.  One can see this more explicitly by writing a standard Dirac spinor as a single column of a $4 \times 4$ matrix where the other columns are set to zero.  Since the gamma matrices are a basis for this matrix, it admits a form like Eq.~(\ref{eq:psi}).

This discussion so far has been in flat space, but the K\"ahler-Dirac equation [Eq.~(\ref{eq:KD})] is valid for any metric, so it applies also in curved space.  This is unlike the usual Dirac equation, which must be generalized by the introduction of a spin connection.  The identification of the K\"ahler-Dirac equation with four copies of the Dirac equation only applies in flat space, but should be approximately true in metrics with small curvature.

A particularly useful feature of the K\"ahler-Dirac formalism is that it extends straightforwardly to triangulated manifolds.  Since the K\"ahler-Dirac formulation contains no spinors, there is no need to introduce a spin-connection or
vielbein.  It is a straightforward matter to apply known results from homology theory \cite{Rabin:1981qj, Montvay:1994cy, Nash:1983cq} to map the exterior derivatives to operators that act on discrete spaces:
\bea  d \mapsto \overline{d}, \\
      \delta \mapsto \overline{\delta}. \ 
\eea
Here $\overline{d}$ is the co-boundary operator and $\overline{\delta}=\overline{d}^T$ is the boundary operator.  These operators act on fields defined on $p$-simplices in a way that is analogous to the action of $d$ and $\delta$ on continuum differential forms.  In a combinatorial triangulation each $p$-simplex is specified by an ordered list of its vertices. The boundary operator $\delta$ acts on such a $p$-simplex to produce a list of its boundary ($p - 1$)-simplices, with elements of the list weighted by signs that denote their orientation within their parent $p$-simplex via the relation
\bea \overline{\delta}_p[v_0,...v_p] = \sum_{i=0}^{p}(-1)^i[v_0,...\hat{v}_i,...v_p],
\eea
where $\hat{v}_i$ indicates that the $i^{\rm th}$ vertex is deleted.  When applied to a $p$-simplex, this operation yields oriented sums of boundary ($p - 1$)-simplex fields for each $p$-simplex.  Conversely, the co-boundary operator applied to a $p$-simplex yields oriented sums of the co-boundary ($p + 1$)-simplex fields that contain that $p$-simplex.

We now describe how the lattice K\"ahler-Dirac matrix is constructed on our triangulated lattices (see \cite{Catterall:2018dns} for the original discussion).  First, an orientation must be chosen for the triangulation.  This is done using a spanning tree, so that when traversing the dual lattice every 4-simplex is visited once.  Second, the vertices of each simplex must be ordered.  Since the vertex labels are all distinct positive integers, we take ascending numerical order as a convenient choice.  The parity of each 4-simplex is then determined by specifying that adjacent faces of adjoining 4-simplices have the opposite parity.  In passing from one simplex to another via the spanning tree we can compare the parity of the neighboring faces (3-simplices) using the boundary operator and the ordering of the vertices of a 4-simplex.  If these faces are of opposite parity, we do nothing, but if they are of the same parity, we flip the signs of all of the faces of the newly visited 4-simplex.  We follow this procedure through the entire spanning tree to completely orient a triangulation.

As an illustration, consider the matrix that has for every 3-simplex the mapping to its oriented co-boundary, that is, an oriented list of 4-simplices.  We denote this matrix $\overline{d}_4$, and its shape is $N_3 \times N_4$. Each column has five nonzero entries, one for each tetrahedron, whose matrix element is either 1 or $-1$, depending on the simplex orientation.  Each row has only two nonzero entries, and one of these is necessarily 1, and the other $-1$, if this
lattice is to be oriented.  

We can continue in this way by constructing $\overline{d}_3$, which is an $N_2 \times N_3$ matrix that has for every 2-simplex the mapping to its co-boundary, and we continue with this same prescription for all of the lower sub-simplex matrices.  One subtlety in our calculation is that we work with degenerate triangulations, where distinct $p$-simplices have the same vertex labels.  In this case, the matrix elements of $\overline{d}_p$ must be specified
using the unique identification label of a simplex, with distinct $p$-simplices treated separately and ignoring the fact that some distinct $p$-simplices share the same vertex labels. 

Putting together these ingredients, the full co-boundary operator is
\bea  \overline{d} = \begin{pmatrix}
                      0 &  0 &  0 &  0 &  0 \\
                      \overline{d}_4 & 0 & 0 & 0 & 0 \\
                      0 & \overline{d}_3 & 0 & 0 & 0\\
                      0 & 0 & \overline{d}_2 & 0 & 0 \\
                      0 & 0 & 0 & \overline{d}_1 & 0 
                      \end{pmatrix}
\eea
where the dimension of this matrix is $N \times N$, with
\bea  N = \sum_{i=0}^{4} N_i.
\eea
The boundary operator is then $\overline{\delta} = \overline{d}^T$, and the lattice K\"ahler-Dirac equation can be written in terms of these as
\bea \label{eq:KDlat}  (\overline{d} - \overline{\delta} + m_0)\omega = 0,
\eea
where it is implicit that $\omega$ is the lattice analog of its continuum counterpart and that $m_0$ multiplies the identity matrix.

The lattice version of the squared K\"ahler-Dirac operator gives the lattice Laplacian
\bea  \Delta = (\overline{d} - \overline{\delta})^2 = -(\overline{d}\overline{\delta} + \overline{\delta} \overline{d}).
\eea
This matrix is block-diagonal, with five blocks, one for each type of $p$-simplex in four dimensions.  Therefore, the square of the K\"ahler-Dirac operator yields lattice Laplacians for not only the
0- and 4-simplices, which have a natural interpretation
in terms of the direct and dual lattices, respectively, but also for all
simplices of intermediate rank.  The square remains block-diagonal when the mass term is added to the kinetic term in the K\"ahler-Dirac operator, a result we make use of in what follows.  Since the blocks result from the square of the K\"ahler-Dirac matrix, each block can be used to source a two-particle bound state.  The diagonal blocks are the easiest to study, and because they include all sub-simplex types, we expect them to be representative of the spectrum.  Thus, they are the focus of this work, as described below.

\section{\label{sec:tasteBreaking} Taste-breaking in two-fermion bound states}

We are interested in the fate of the flavor-symmetry of K\"ahler-Dirac fermions when they live on the dynamical geometries of EDT.  If the lattice fermions are to behave as in the continuum, in the flat-space limit we expect that a single flavor of K\"ahler-Dirac fermion will reduce to four identical copies of Dirac fermions.  However, even in the continuum, the K\"ahler-Dirac action coupled to geometry breaks the flavor symmetry, hereafter referred to as taste-symmetry, so that the different taste fermions couple differently to gravity.  We thus do not expect the various two-particle bound states of the different taste fermions to be degenerate, but it is an open question as to which scale sets the non-degeneracy.  

K\"ahler-Dirac fermions have an exact $U(1)$ symmetry that generalizes the $U(1)$ symmetry of staggered fermions in flat space lattice field theory, and one of its consequences is that it ensures that the bound states associated with the $p_4$ and $p_0$ blocks remain massless in the chiral limit.  It was also shown in Ref.~\cite{Catterall:2018dns} that the chiral condensate vanishes in the large volume limit, providing further evidence that the $U(1)$ symmetry is not spontaneously broken at the Planck scale in the EDT formulation.  (Similar results that show an absence of chiral symmetry breaking for fermions coupled gravitationally have been found in different scenarios using the FRG; see \cite{Eichhorn:2011pc, Meibohm:2016mkp, Eichhorn:2018yfc, deBrito:2020dta, deBrito:2023zxl, deBrito:2023kow, Eichhorn:2026euv}.)  Although the simulation values of the $p_4$ and $p_0$ bound-state masses appear to remain massless in the chiral limit, the other taste combinations are not protected, and we therefore expect that their corresponding bound-state masses will not vanish in the chiral limit.  The question then is what scale sets the value of the bound-state masses in the chiral limit?  One candidate is the lattice cutoff scale (inverse lattice spacing), where the bound states not protected by the $U(1)$ symmetry acquire masses of order the cut-off, and thus decouple from the theory.  Another possibility is that the bound states acquire masses of order the Planck scale.  Yet another possibility is that they acquire masses of order the Hubble scale, in which case the taste-breaking scale is that of the curvature scale of the background geometry.  
A final possibility is that the bound states in the chiral limit acquire masses that scale as a positive power of the lattice spacing, vanishing in the continuum limit.  This would be the gravity analogue closest to that of staggered fermions in QCD, in which case the K\"ahler-Dirac and (four copies of) Dirac fermions would be in the same universality class even in the presence of gravity.  Although the possibility that the masses become of order the cut-off and decouple from the theory was disfavored in Ref.~\cite{Catterall:2018dns}, the other possibilities were not ruled out, especially in light of the large systematic errors later found in the relative lattice spacings used in that work.  We revisit the taste-breaking of the two-particle bound states of K\"ahler-Dirac fermions here, taking advantage of the finer lattice spacings and more reliable determinations of those spacings now available \cite{Dai:2024vjc}.

\subsection{Computing fermion correlation functions}

%
%
%
%

We calculate the fermion bound-state masses using two-point correlation functions, following the construction in \cite{Catterall:2018dns}.  The correlation functions are of the form,
\bea  \langle \overline{\omega}_x \Gamma \omega_x \overline{\omega}_y \Gamma \omega_y \rangle &\rightarrow 
     \langle (\omega_y \overline{\omega}_x)(\omega_y \overline{\omega}_x)^\dagger \rangle \ \ \ \ \ \ \ \ \ \ \ \ \ \ \   \\
\label{eq:twoprops}     &= \langle(\overline{K}+m_0)^{-1}_{yx} ((\overline{K}+m_0)^{-1}_{yx})^*\rangle,  \\
     &\equiv \langle \pi(x)\pi(y) \rangle, \ \ \ \ \ \ \ \ \ \ \ \ \ \ \ \ \ \ \ \ \ \ \ 
\eea
where we set $\Gamma =1$ to pick out the scalar bound state.  The inverse of the K\"ahler-Dirac propagator can be evaluated by noting that the product of two inverse matrices is the inverse of the product, $(AB)^{-1}=B^{-1} A^{-1}$.  The product of the two propagators in Eq.~(\ref{eq:twoprops}) is $\overline{K}\overline{K}^T + m_0^2$, which is the block-diagonal Klein-Gordon propagator.  Note that the inverse of a block diagonal matrix is itself block diagonal.  Then, if the source and sink for the correlation function live in the same block for both K\"ahler-Dirac propagators, then we only need the inverse of each block of the Klein-Gordan matrix individually.  Each of the five blocks corresponds to one of the types of sub-simplex $p_i$, from the point vertex ($i=0$) up to the four-simplex ($i=4$).  

It is convenient to use the dual (simplex) geodesic distance for our correlator source-sink separation.  This is straightforward when our source and sink are both drawn from the $p_4$ block, since each source and sink is a single 4-simplex.  In this case we average the sink over all 4-simplices a common geodesic distance away from the source 4-simplex.  When the source is drawn from a sub-simplex block ($i<4$), we smear the source over a single 4-simplex so that, for example, a $p_2$ source is uniformly weighted over the 10 triangles of a randomly chosen 4-simplex.  We use the same smearing for the sink, which we then average over all sink 4-simplices that are the same geodesic distance from the source.

This construction leads to five distinct taste bound states, which are the easiest to construct.  There are others that can be constructed from non-block diagonal propagators, but we expect that the diagonal ones are representative of the taste-breaking of the system, and we restrict our study to these five.  

All source simplices are selected randomly from the largest three-volume cross-section of the entire lattice. This is done by first shelling a configuration starting from a source chosen at random, and then only selecting sources for the propagators from the largest slice in the shelling. We find that restricting our sources to come from the largest three-slice minimizes finite lattice spacing effects, and it is the same procedure that has been implemented in previous work on EDT \cite{Laiho:2016nlp, Catterall:2018dns, Dai:2021fqb}.  We use 60 sources per configuration, chosen at random without replacement, except for our noisiest correlators ($p_1$, $p_2$, and $p_3$) on a single ensemble at our finest lattice spacing ($N_4$=32k and $\kappa_2=3.8$), where we use 300 sources per configuration (again chosen without replacement).  We perform a direct sparse solve using lower-upper (LU) decomposition for our propagator inversions, except in the case of the $p_1$ inversions at some of our larger volumes, because our off-the-shelf direct solver requires too much memory.  In those cases we instead use a biconjugate gradient stabilized inverter for computing the propagators.  This takes much longer than a typical direct solve at this volume, but does not suffer from memory issues.  


\subsection{Extracting bound-state masses}

Assuming a massive ground state dominates for large source-sink separation, we expect the two-point correlation functions to take the asymptotic form \cite{deBakker:1996qf}
\bea \label{eq:propagator_fit}
\langle \pi(x)\pi(y) \rangle \simeq a \exp{(-m|x-y|)}/|x-y|^b,
\eea
where $a$, $b$, and $m$ are free parameters in the fit, with $m$ the mass of the fermion bound-state.


\begin{figure*}[tb] 
    \centering
    \includegraphics[width=\textwidth, height=0.9\textheight, keepaspectratio]{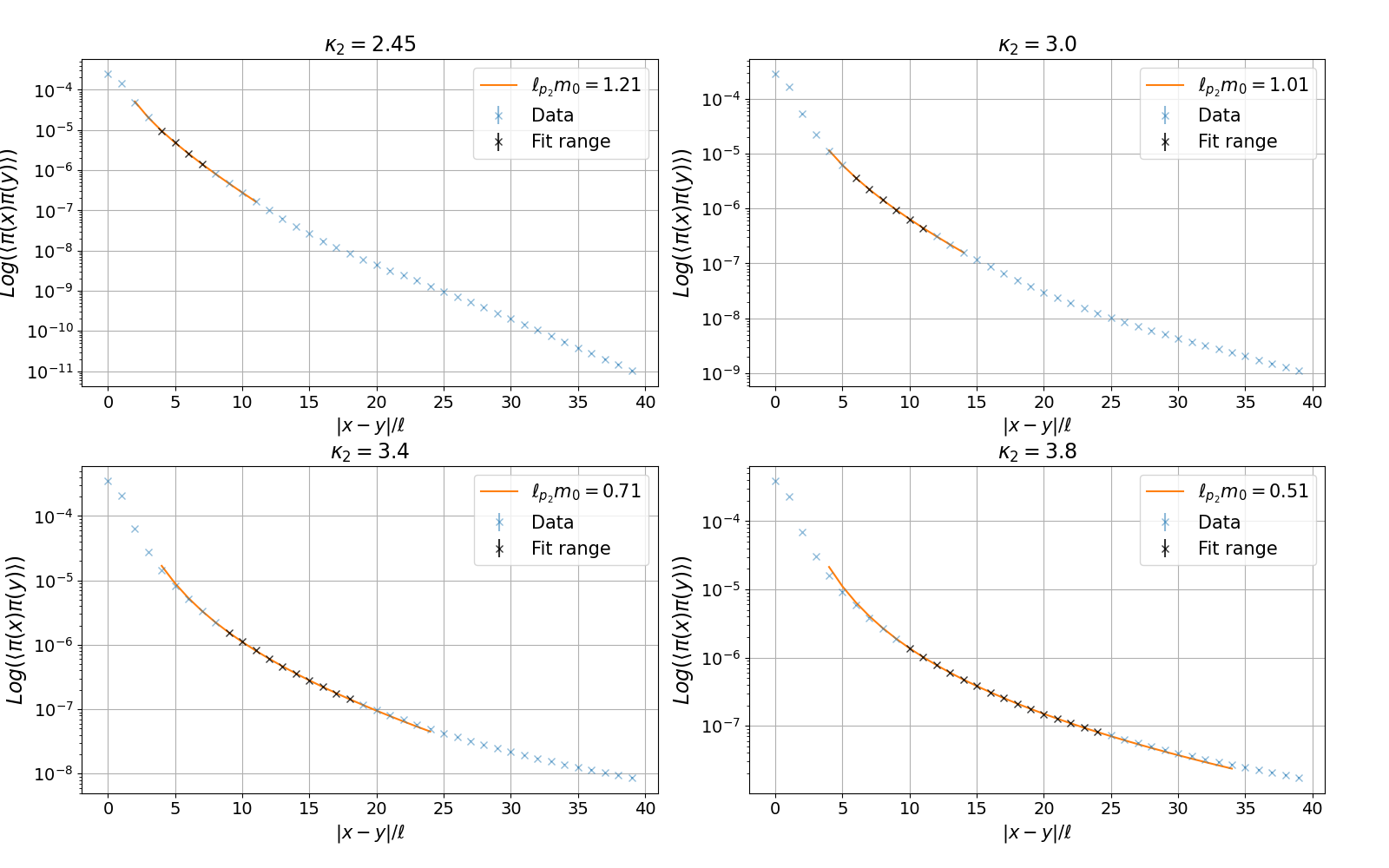}
    \caption{Sample correlation function fits for the $p_2$ block fermion bound state.  The logarithm of the two-point correlation function $\langle \pi(x)\pi(y) \rangle$ is shown versus the Euclidean separation of the source and sink, $|x-y|/\ell$, which is expressed in $\ell$ units.  Statistical error bars are smaller than plot symbols.  Each plot shows a fit at a different lattice spacing, where the fit window (expressed in $\ell$ units) and the bare mass used to compute the propagator (expressed in $\ell_{p_2}$ units) have been rescaled in lattice units to roughly match in physical units across all four plots.  The fit to Eq.~(\ref{eq:propagator_fit}) is shown in orange, and the points included in the fit are highlighted in black. }
    \label{fig:corr_fits}
\end{figure*}

We are making some additional assumptions when using Eq.~(\ref{eq:propagator_fit}) for our analysis, since this formula is only valid for a massive scalar propagating on a flat space background in infinite volume.  In order to be applicable, the minimum source-sink separation must not be too small, since at short Euclidean distances the correlators will be contaminated by excited states and, for sufficiently short distances, by lattice discretization effects.  The source-sink separation must also not be too large, since at long distances the de Sitter curvature becomes important and Eq.~(\ref{eq:propagator_fit}) no longer applies.  Thus, we would like to restrict our fit ranges to values much smaller than the de Sitter radius, though in practice this is difficult to implement, since our lattice volumes are not very large.  The maxima of our source-sink separations for our central fit ranges are around two to three times smaller than the de Sitter radii of our lattices, while the minima are around four to six times smaller.  Another concern in applying Eq.~(\ref{eq:propagator_fit}) is that at long distances, there are elongated structures that branch out from the mother universe in our geometries, leading to deviations from the classical de Sitter solution \cite{Laiho:2016nlp}, though the deviations get smaller as the continuum limit is approached.  Despite our attempts to work at sufficiently small source-sink separations to avoid these issues, we expect that the lack of flatness in our geometries will lead to deviations from the flat-space expectation.  

This brings us to another consideration when using Eq.~(\ref{eq:propagator_fit}) in our analysis, namely the role that the power-law contribution plays.  In the infinite-volume, flat-space limit of the free-field scalar propagator, we know that $b=2$.  Here, however, we might not expect to recover this result, since we are summing our sink over a spherical spatial slice.  Instead we may pick out the lowest spherical harmonic as the ground state, with the expectation that there will be gaps in the low-lying energy levels due to the finite volume of the geometry.  In that case, the continuous momentum integral would become a discrete sum, which for small volumes would not be sufficiently densely spaced to give rise to the power law in Eq.~(\ref{eq:propagator_fit}).  Thus, the role of $b$ in our fits is that of a nuisance parameter, which allows for deviations from flat space, either from the constant curvature of the de Sitter geometry or the discretization effects that cause deviations from de Sitter at long distances on our lattices.  We discuss our findings for the behavior of $b$ in our analysis below.

We choose our fit range so that the fit ansatz Eq.~(\ref{eq:propagator_fit}) is a good description of the data, where the precise choice of this fit range is motivated by the considerations discussed above.  The fit window is scaled roughly by 1/$\ell_{rel}$ from our fiducial lattice spacing ($\kappa_{2} =$ 3.0) so that the fit ranges have similar values in physical units at each lattice spacing.  In this rescaling we favor larger minimum source-sink separations for our central fits at coarser lattice spacings, since we want to avoid the discretization errors that we expect to be present when the source and sink are only a few lattice spacings apart.  The central fit ranges that we choose are given in Table~\ref{table:fitranges}, and a sample of our fits to correlation functions is shown in Fig. (\ref{fig:corr_fits}). 

\begin{table}[htbp]
\centering
\caption{Fit ranges used for the propagator fits, including those shown in Fig.~\ref{fig:corr_fits}. The parameter $\kappa_2$ labels the lattice spacing.}
\label{table:fitranges}
\small
\setlength{\tabcolsep}{12pt} 
\begin{tabular}{cc} 
\toprule
$\kappa_2$ & Fit window \\
\midrule
2.45 & $|x-y|/\ell \in [4,8]$\\
3.0  & $|x-y|/\ell \in [6,12] $\\
3.4  & $|x-y|/\ell \in [9,19] $\\
3.8  & $|x-y|/\ell \in [10,25] $\\
\bottomrule
\end{tabular}
\end{table}


Figure~\ref{fig:corr_fits} shows fits across all four lattice spacings for the $p_2$ (triangle) block of correlators, where the fit range has been adjusted to correspond to roughly the same window in physical units.  The $p_2$ correlators are representative of the other blocks, and these plots are sufficient to illustrate the features of our correlator fits.  The masses in the plots have been adjusted to match in physical units.  Note that the bare mass of each $p_i$ block is expressed in the lattice units corresponding to the distance step of the covariant Laplacian of the $p_2$ block, and these units must be converted to $\ell$ units as part of our analysis.  We denote the lattice units of each $p_i$ block as $\ell_{p_i}$, and the determination of these spacings is discussed in more detail in the following subsection.  The bare masses of Fig.~\ref{fig:corr_fits} are expressed in $\ell_{p_2}$ units.  We plot the logarithms of the correlators, which for pure exponential fall off should be straight lines.  However, the power law term in Eq.~\ref{eq:propagator_fit} obscures this picture somewhat, leading to non-vanishing curvature.  We see in the upper left plot of Fig.~\ref{fig:corr_fits}, which is our coarsest lattice spacing, that for sufficiently large source-sink separation the concavity changes sign, an effect that is not accommodated by our fit ansatz, Eq.~(\ref{eq:propagator_fit}), and is not expected for a background of constant curvature like de Sitter.  This shift in concavity is likely an artifact from probing too large of a source-sink separation, where the correlator is sensitive to the elongated structures that appear in our lattice geometries.  Note that this change in concavity gets pushed to source-sink separations that are larger in lattice units as we make the lattice spacing smaller.  Figure~\ref{fig:corr_fits} shows fits to Eq.~(\ref{eq:propagator_fit}), where the data points used in the fits are highlighted in black.  Statistical errors are smaller than the symbols for the data points shown in the plots.

\begin{figure*}[htb]
\begin{minipage}{0.475\linewidth}
    \centering
    \includegraphics[width=\linewidth]{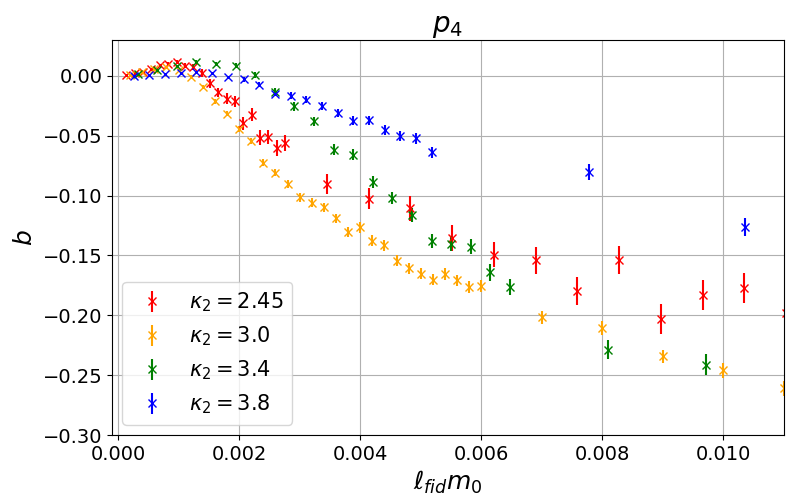}
\end{minipage}
\begin{minipage}{0.475\linewidth}
    \centering
    \includegraphics[width=\linewidth]{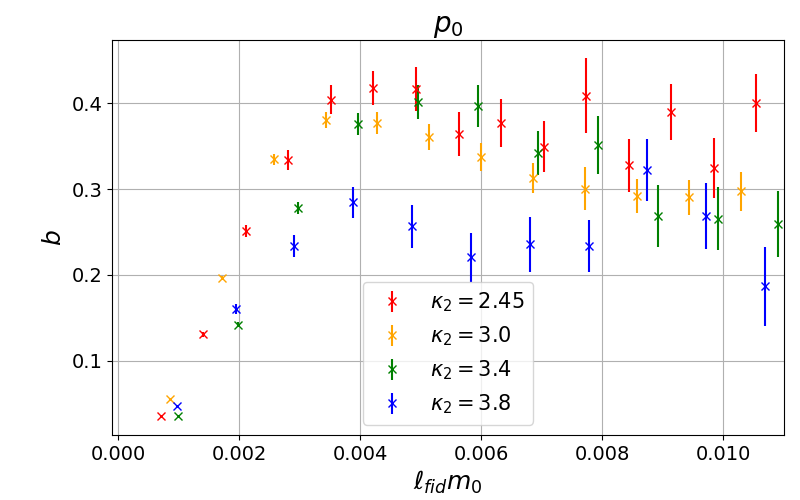}
\end{minipage}
\begin{minipage}{0.475\linewidth}
    \centering
    \includegraphics[width=\linewidth]{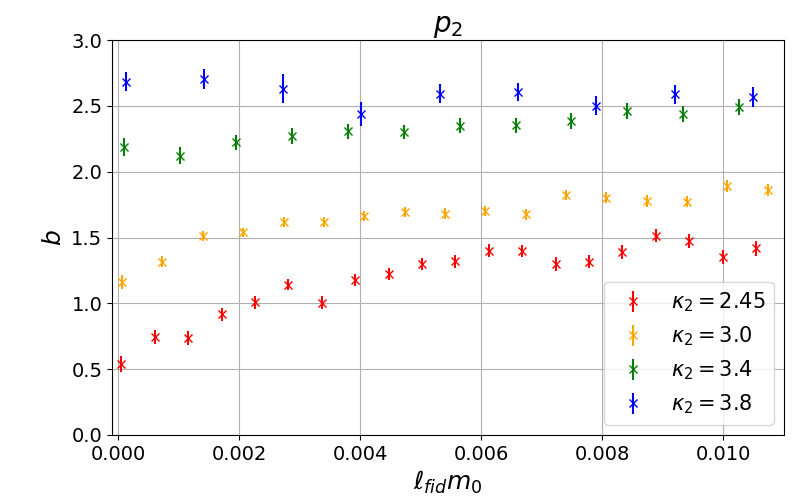}
\end{minipage}
\caption{The exponent $b$ obtained from fits to Eq.~(\ref{eq:propagator_fit}) versus the input bare fermion mass.  The bare fermion masses have been rescaled to put them in common units across different $p_i$ blocks and lattice spacings.  The trend in $b$ for the $p_4$ and $p_0$ blocks is towards zero as the continuum limit is approached.  For the $p_2$ block, $b$ continues to grow as the lattice spacing decreases, but more slowly on our finest lattice spacings. \newline
} 
\label{fig:b}
\end{figure*}

We investigate the behavior of the exponent $b$ in Eq.~(\ref{eq:propagator_fit}).  Although we find only modest dependence of $b$ on volume, we find fairly strong dependence on lattice spacing, as can be seen in Fig.~\ref{fig:b}. In these plots the exponent $b$ is plotted against the bare fermion mass, which has been rescaled to the same physical units for each lattice spacing using the prescription discussed in more detail in the following subsection.  For the $p_4$ and $p_0$ blocks, it appears that $b$ gets smaller across the mass range as the lattice spacing is decreased.  However, for $p_4$ this decrease is not monotonic, as $b$ first increases as the lattice spacing decreases, only to drop significantly at our finest lattice spacing.  Based on this, we conclude that the power-law dependence in our correlators is dominated by discretization effects, which look like they are trending towards zero in the continuum limit for the $p_4$ and $p_0$ blocks.  For the other three blocks, $b$ increases as the lattice spacing decreases, but more slowly as we go to finer lattice spacings.  Fig.~\ref{fig:b} shows this for the $p_2$ block.  The question as to whether the $b$ values for the $p_1$, $p_2$, and $p_3$ blocks eventually begin to decrease towards zero at still finer lattice spacings will have to await future studies.  A zero value for $b$ in the continuum limit is consistent with the expectation of a gap in the spectrum for the scalar states, as seen for the $p_4$ and $p_0$ blocks.  We continue to work under the assumption that including the power-law dependence in Eq.~(\ref{eq:propagator_fit}) is an adequate model for extracting ground-state masses from the correlation functions for all of the $p_i$, with the expectation that this interpretation becomes more accurate at finer lattice spacings.

The fits to Eq.~(\ref{eq:propagator_fit}) are done using non-linear least squares fitting including the correlations of the dependent data. The errors are estimated using single-elimination jackknife resampling, including the off-diagonal terms in the correlation matrix.  We estimate the size of autocorrelation errors using a blocking procedure.  The data is blocked until the errors no longer increase.  The error versus block size is rather noisy, so we automate a procedure that roughly locates the mean of the plateau as a function of block size where the errors stop increasing.  In order to retain enough information to resolve the correlation matrix when performing fits, the data is not blocked, but the errors are inflated to reflect the increased error due to autocorrelations. The fits are performed under a jackknife, and the correlation matrix is reconstructed for each individual fit under the jackknife from the data on each jackknife subensemble. Since we include correlations in the fit, we expect the $\chi^2$ per degree of freedom to be a reliable measure of the goodness of fit. We compute from the $\chi^2$ and the number of degrees of freedom a confidence interval (a $p$-value) for the fit, correcting for finite sample size \cite{Toussaint:2008ke}.  

\begin{figure}
    \centering
    \includegraphics[width=1.0\linewidth]{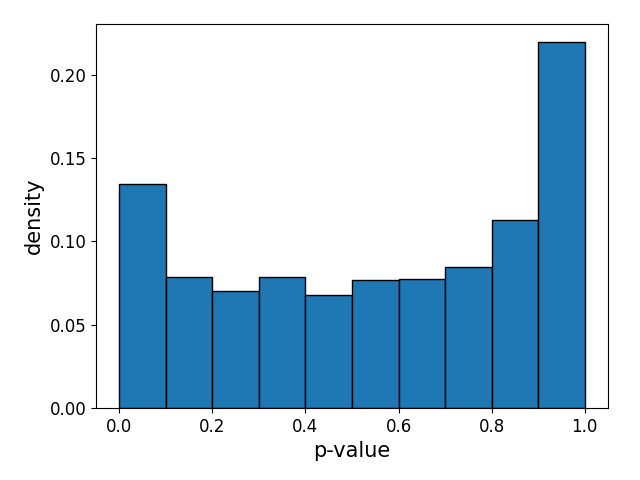}
    \caption{A normalized histogram of $p$-values for all of the correlator fits across every ensemble, each of the five $p_i$ blocks, and every mass value considered in this analysis.}
    \label{fig:p-vals}
\end{figure}

A histogram of $p$-values from our central fits is shown in Fig.~\ref{fig:p-vals}.  This histogram includes fits for all taste-bound states across all of our volumes and four lattice spacings, leading to around 2000 separate fits.  The resulting histogram is expected to be uniform, and we see that although it is close to uniform, there are spikes in the highest and lowest bins.  The spike in the lowest bin ($p$-value<0.1) is mostly due to fits to the $p_4$ bound state.  The $p_4$ state is the least noisy, so Eq.~(\ref{eq:propagator_fit}) may not be sufficient to describe the data over the whole range of masses and ensembles.  The spike at small $p$-values suggests that the $p_4$ errors may be slightly underestimated, but for the purposes of characterizing the behavior of the taste-splittings, this is not a big effect.  The spike at high $p$-values ($p$-value >0.9) is mainly due to the fits to the $p_2$ bound states, suggesting that the errors in the $p_2$ state are somewhat overestimated.  Most likely we would need longer runs to enable even better estimations of the autocorrelation errors, especially on our largest volumes, in order to get a flat $p$-value distribution.  For our purposes, this distribution is sufficiently flat to validate the use of Eq.~(\ref{eq:propagator_fit}) as a good description of our correlator data.

A fit systematic error is estimated by varying the minimum and maximum values of the fit range by some fixed amount and taking the larger difference between those two variant fits from the central value as a systematic error.  For the two coarsest lattice spacings ($\kappa_2=2.45$ and 3.0) the variation is plus or minus one lattice step and for the two finer lattice spacings ($\kappa_2=3.4$ and 3.8) the variation is plus or minus two steps.  The error quoted for the bound-state mass includes this systematic error added in quadrature with the statistical error of the fit. 


\subsection{Chiral extrapolations}

\begin{figure*}[tb]
    \centering
    \includegraphics[width=\textwidth, height=0.9\textheight, keepaspectratio]{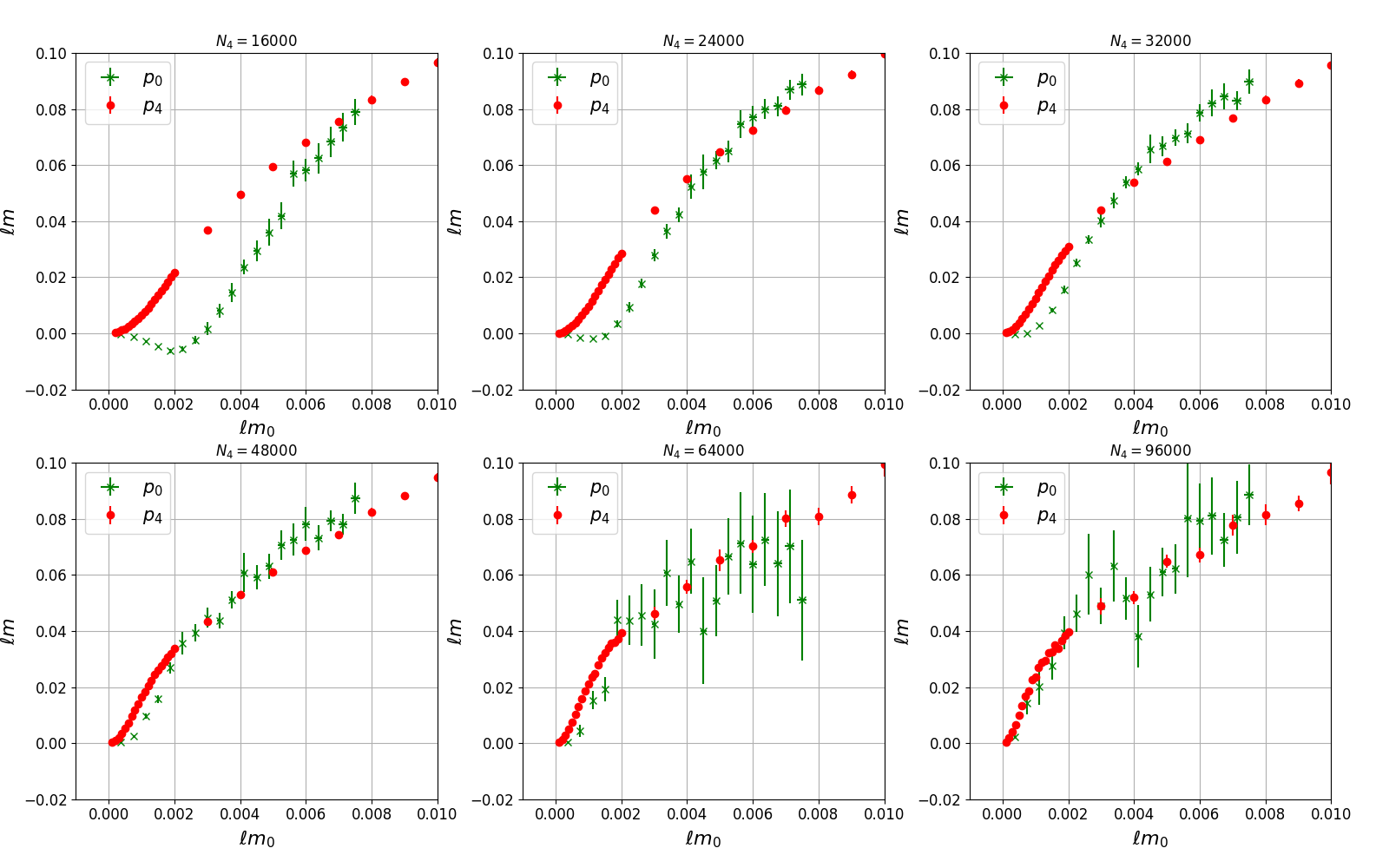}
    \caption{Chiral extrapolation data for the $p_{0}$ and $p_{4}$ bound state masses on every volume of our our finest lattice spacing at $\kappa_2$ = 3.8.  Each plot shows the two-particle bound state masses versus the bare fermion masses, all expressed in $\ell$ units.  Agreement between the $p_4$ and $p_0$ data improves as the volume increases.}
    \label{fig:quad_and_point}
\end{figure*}

We are interested in the behavior of the bound-state masses extracted from our correlation function fits as the bare fermion masses approach zero, i.e. in the chiral limit.  The $U(1)$ symmetry of K\"ahler-Dirac fermions is a remnant of the full chiral symmetry, and it ensures that a subset of the bound-states (those from the $p_4$ and $p_0$ blocks) vanish as the chiral limit is approached.  Figure~\ref{fig:quad_and_point} shows the $p_4$ and $p_0$ bound-state masses as a function of the bare fermion mass on a series of volumes at our finest lattice spacing ($\kappa_2=3.8$).  These plots illustrate a few features of the calculation.  First, the bound-state masses for the $p_4$ and $p_0$ blocks do vanish as the bare fermion masses go to zero, as expected if the $U(1)$ symmetry is unbroken. Second, the finite-volume effects are different for different taste bound states, but the results appear to converge for large volumes.  This happens at sufficiently fine lattice spacings, though it requires fairly large volumes, as seen here.  

Third, although the mass of the bound state $m$ inherits the renormalized dual lattice units $\ell$ of the Euclidean source-sink separation, $m_0$ inherits the units of the covariant Laplacian of the (Klein-Gordon) kinetic term.  The $p_4$ kinetic term is by its definition formulated on the dual lattice, but the $p_0$ kinetic term is formulated on the direct lattice.  Thus, we might expect the bare masses of these two different blocks to be related by the (renormalized) direct to dual lattice spacing ratio $a_{\rm lat}/\ell$.  This rescaling has been performed in Fig.~\ref{fig:quad_and_point} to convert the $p_0$ mass to dual ($\ell$) units.  The $a_{\rm lat}/\ell$ conversion factors used here were determined in Ref.~\cite{Dai:2024vjc} by demanding that the geometries be consistent with the classical de Sitter solution in the large-volume, continuum limit.  They emerge from that analysis because the shelling function used to determine the shape of the lattices is naturally formulated in terms of both the direct and (renormalized) dual spacings, and demanding that the geometries approach the de Sitter solution fixes the conversion factor between them.  (The error in the determination of $a_{\rm lat}/\ell$ leads to a small horizontal error on the rescaled $p_0$ block mass, barely visible in Fig.~\ref{fig:quad_and_point}.)  The fact that the chiral extrapolations of the bound state masses of the $p_4$ and $p_0$ blocks agree when the bare mass of the $p_0$ state is rescaled by this factor (in the large volume limit) provides independent evidence that this rescaling factor is correctly determined and that therefore the geometry emerging from our simulations is indeed de Sitter.

This comparison of the $p_4$ and $p_0$ bound state masses as a function of bare fermion mass illustrates the need for the rescaling of the bare fermion masses in the different blocks of the K\"ahler-Dirac matrix.  The $p_4$ block is formulated directly in $\ell$ units, but the other four blocks require rescaling factors, denoted generically $\ell_{p_i}/\ell$.  The $p_0$ rescaling factor may be obtained from $a_{\rm lat}/\ell$ as we have seen, but what of the remaining three?  The procedure we adopt is as follows.  We match the shapes of the bound-state masses $\ell m$ as a function of the bare fermion masses $\ell_{p_i}m_0$ to that of the $p_4$ curve in the large-volume limit by adjusting the factor $\ell_{p_i}/\ell$ until the curves overlap.  This is complicated by the fact that the bound states of the $p_1$, $p_2$, and $p_3$ blocks do not vanish in the chiral limit, at least not at finite lattice spacing.  We therefore match the shape of the chiral extrapolation curves up to a non-zero vertical offset in the chiral limit.  This offset is very similar to what occurs in the pion sector of Lattice QCD with staggered fermions, where it is known as a ``staggered taste-splitting.''  We adopt that terminology here.   

\begin{figure*}[p] 
    \centering
    \includegraphics[width=\textwidth, height=0.9\textheight, keepaspectratio]{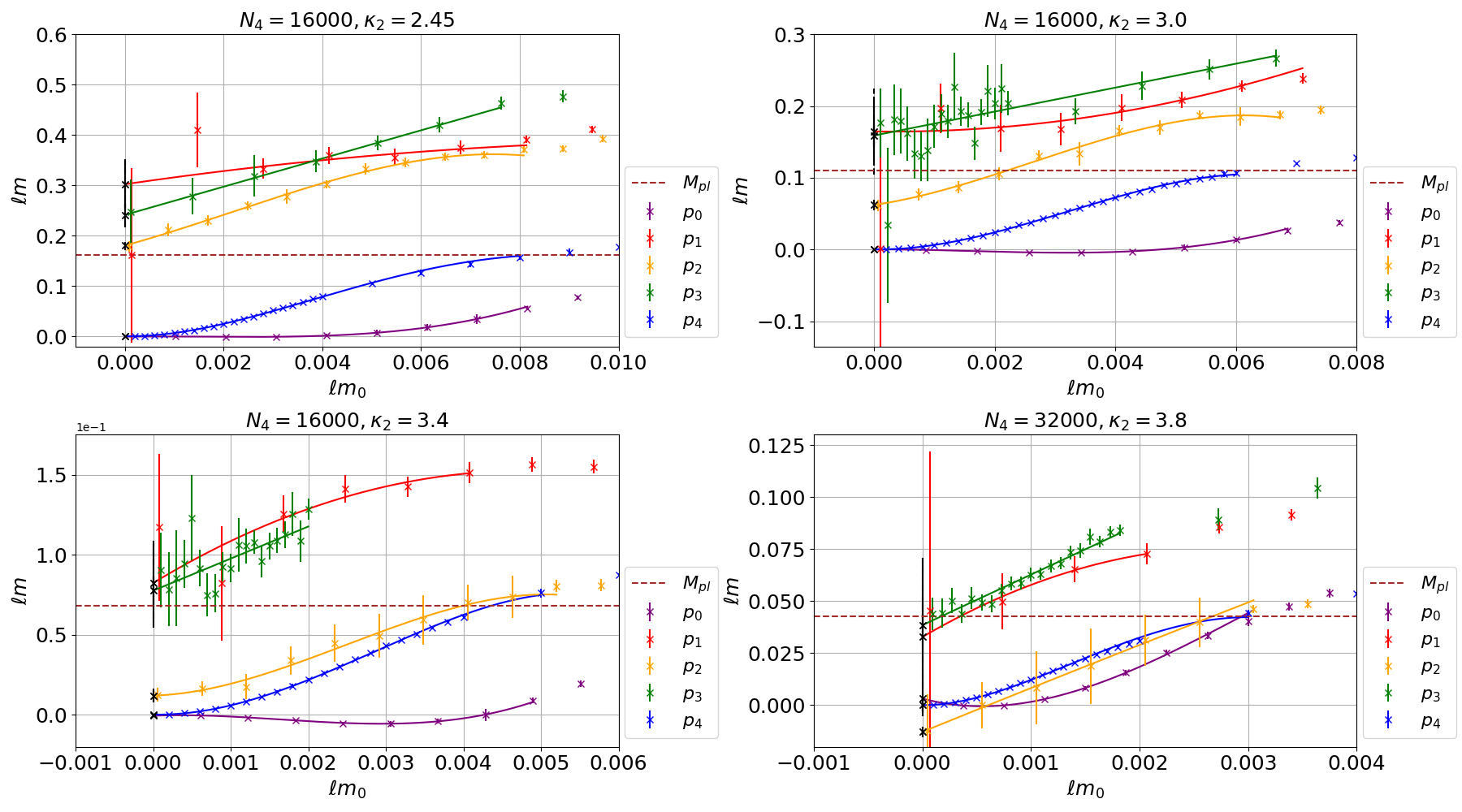}
    \caption{Sample chiral extrapolations at each lattice spacing.  The bound-state masses are plotted versus the bare fermion masses for each of the five $p_i$ blocks.  All masses are (re)expressed in $\ell$ units.  Polynomial fit functions used to extrapolate the bound state masses to the chiral limit are shown.  Extrapolated values with statistical errors (solid) are shown in black, and the total error (dashed) includes a chiral fit systematatic error added in quadrature to the statistical error.  In all cases the $p_0$ and $p_4$ states extrapolate to zero, as these sectors are protected by a $U(1)$ symmetry, which is a remnant of chiral symmetry.  The central value for the Planck mass (as determined on these ensembles in \cite{Dai:2024vjc}) at each lattice spacing is shown as a dotted horizontal line for scale. }
    \label{fig:chiral_fits}
\end{figure*}

Figure~\ref{fig:chiral_fits} shows the chiral extrapolations of all five taste bound-states at representative volumes on our four lattice spacings.  The different $p_i$ bound-state masses are plotted versus the bare fermion masses, where the bare fermion masses have been rescaled from $\ell_{p_i}$ units to $\ell$ units according to the prescription described above, and the rescaling factors $\ell_{p_i}/\ell$ determined from this procedure are given in Tab.~\ref{tab:rescaling_factors}.  We do not quote errors for the conversion factors from $\ell_{p_1}$, $\ell_{p_2}$, and $\ell_{p_3}$ units, since in this work they are mainly used for presentation purposes, e.g. rescaling the horizontal axes of the data sets appearing in the plots of Fig.~\ref{fig:chiral_fits}.  Fits to the data are performed in order to extract the chiral limits of the bound-state masses; a sample of these fits is shown in Fig.~\ref{fig:chiral_fits}, and they are described in more detail below.  It can be seen in the plots that the $p_4$ bound states have the smallest errors, while the $p_3$ and $p_1$ states are rather noisy.  Also shown for reference in the plots of Fig.~\ref{fig:chiral_fits} is a dashed line at the Planck mass in $\ell$ units at that lattice spacing.

\begin{center}
\begin{table}
\caption{Scaling factors for bare quark masses.  The first column is $\kappa_2$, which labels the lattice spacing.  The second column labels the $p_i$ block, and the third column is the ratio $\ell_{p_i}/\ell$ needed to convert the bare mass in the $p_i$ block into $\ell$ units.  Note that $\ell_{p_4}\equiv\ell$.}
\label{tab:rescaling_factors}
\begin{tabular}{c c c} 
\hline \hline
$\kappa_2$& $p_{i}$ & $\ell_{p_i}/\ell$ \\
\hline
\multirow{3}{4em} {\centering 2.45}& $p_{0}$ & 9.81 \\
& $p_{1}$ & 75\\
& $p_{2}$ & 125 \\
& $p_{3}$ & 8\\ 

\hline
\multirow{3}{4em} {\centering 3.0}& $p_{0}$ & 11.67 \\
& $p_{1}$ & 100\\
& $p_{2}$ & 150\\
& $p_{3}$ & 9\\
\hline
\multirow{3}{4em} {\centering 3.4}& $p_{0}$ & 16.34 \\
& $p_{1}$ & 125\\
& $p_{2}$ & 175\\
& $p_{3}$ & 10\\
\hline
\multirow{3}{4em} {\centering 3.8}& $p_{0}$ & 26.67 \\
& $p_{1}$ & 150\\
& $p_{2}$ & 200\\
& $p_{3}$ & 11\\
\hline\hline
\end{tabular}
\end{table}
\end{center}

The chiral fit functions are taken to be polynomial in the bare mass, where the order of the polynomial is chosen to be large enough to describe the data.  (The polynomials we use are cubic for $p_{0}$ and $p_{2}$, quadratic for $p_{1}$, linear for $p_{3}$, and quartic for $p_{4}$.)  We use these fits to extract the values of the bound-state masses in the chiral limit, i.e. the taste-splittings.  For the $p_4$ bound-state mass, the fit range for $\ell m_0$ is chosen to be between zero and the value of $\ell m_0$ corresponding to the Planck mass.  Chiral fits for the other $p_i$ are also taken roughly within this same range in $\ell m_0$, even though for some of the $p_i$, because of the taste-splittings, the bound-state masses $m$ exceed the Planck mass.  As can be seen in Fig.~\ref{fig:chiral_fits}, the non-zero taste-splittings appear to get smaller in physical units as we go to finer lattice spacings, a statement we make more precise in the following subsections.  

\begin{figure}
    \centering
    \includegraphics[width=1.0\linewidth]{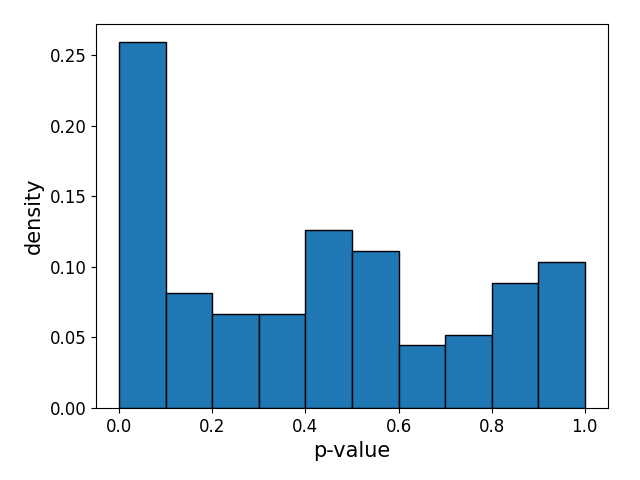}
    \caption{The normalized $p$-value distribution including every chiral extrapolation fit for all ensembles and all five $p_i$ blocks.}
    \label{fig:chiral_pvals}
\end{figure}

The masses obtained from the correlator fits are passed to the chiral fitter under a jackknife, and as in the case of the correlator fits, the chiral fits are performed under a jackknife, where the correlation matrix is reconstructed for each individual fit under the jackknife from the data on each jackknife subensemble.  Once again, we compute from the $\chi^2$ and the number of degrees of freedom a $p$-value for each fit, correcting for finite sample size as in \cite{Toussaint:2008ke}.  Figure~\ref{fig:chiral_pvals} shows a histogram of $p$-values from all of the chiral fits for all of the $p_i$ bound-state masses on all of our volumes and lattice spacings.  The distribution of $p$-values here is fairly flat, with a spike in the lowest bin.  This spike is mainly due to the $p_4$ block, which has the smallest statistical errors of any of the blocks.  Again, the $p$-value distribution suggests that the errors of the $p_4$ block may be somewhat underestimated.  We could include an error inflation factor to account for this, but this would only increase the already very small $p_4$ errors by a factor of ${\cal O}(1)$, which would not change our conclusions.  

In estimating the error for the taste-splittings in the chiral limit, we include a fit systematic error added in quadrature with the jackknife statistical error.  This fit systematic error is determined by looking at the difference between the central fit and fits with one and two successively larger masses included and taking the largest difference from the central fit as a systematic error.  In Fig.~\ref{fig:chiral_fits} the statistical errors on the extrapolated values are shown as solid riser lines, and the total errors including the systematic errors are shown as dashed riser lines.

\subsection{Finite-size effects}

\begin{figure*}[p] 
    \centering
    \includegraphics[width=\textwidth, height=0.9\textheight, keepaspectratio]{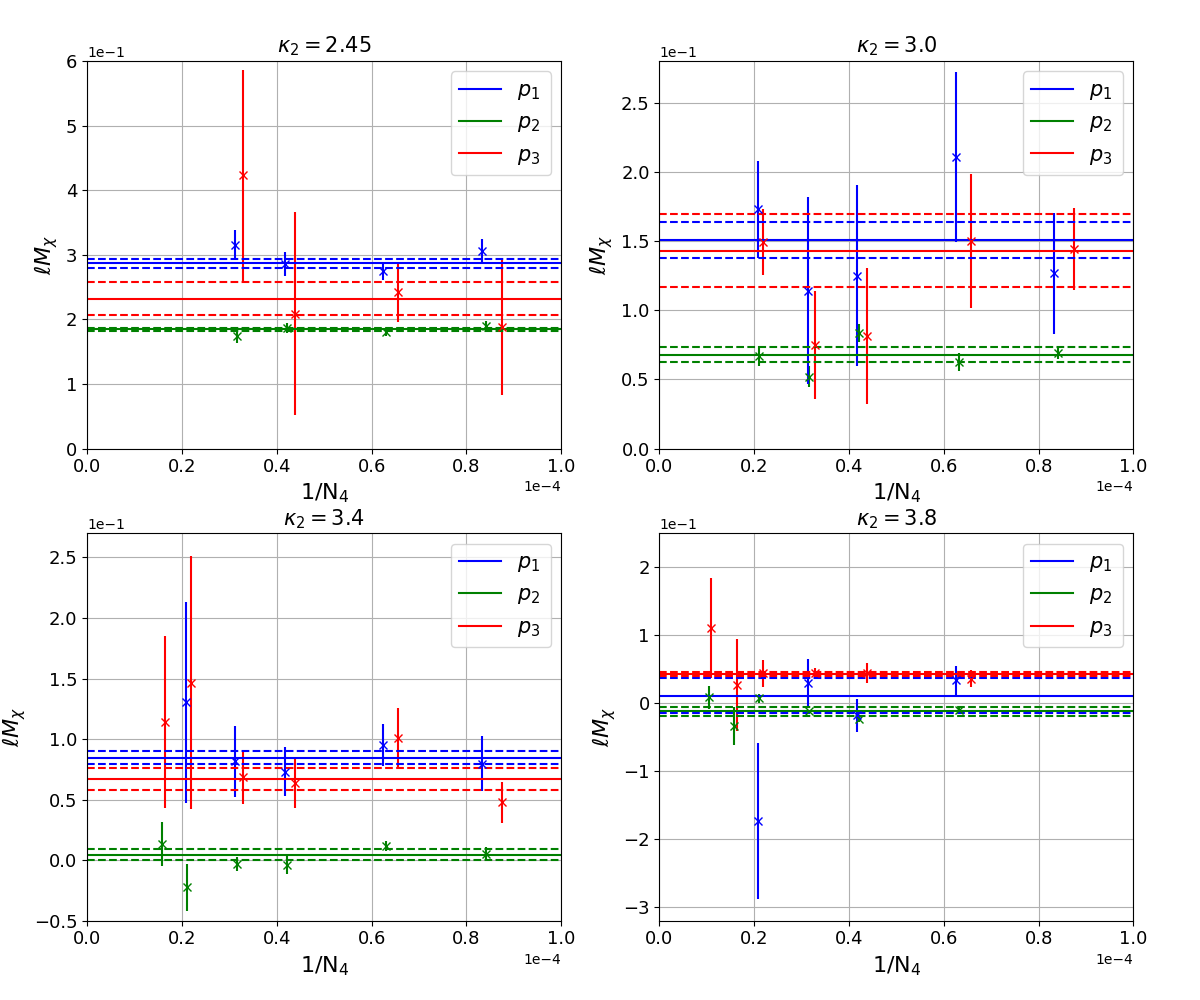}
    \caption{Taste-splittings in the chiral limit plotted versus the inverse volume at each lattice spacing, including fits to a constant and the corresponding $1\sigma$ error bars. Small horizontal offsets have been introduced for the $p_{2}$ and $p_{3}$ taste-splittings for clarity. }
    \label{fig:Bandplots}
\end{figure*}

In the remainder of this work we focus on the non-zero taste-splittings $M_\chi$, which are the chiral limits of the taste bound states corresponding to $p_i$ blocks with $i=1$, 2, and 3. Figure~\ref{fig:Bandplots} shows a finite-volume study, with all of the $M_\chi$ values (in $\ell$ units) plotted as a function of $1/N_4$, at each of our four lattice spacings.  Looking at the behavior of the taste-splittings as a function of lattice volume, there is no clear trend as the volume is varied.  As Fig.~\ref{fig:Bandplots} also shows, the $p_1$ and $p_3$ splittings are quite noisy.  Given the very mild dependence of the splittings on volume, we fit the $M_\chi$ dependence on volume to a constant, and we take these results to be representative of the non-zero taste-splittings at a given lattice spacing.  These final values are given in Table~\ref{tab:band_fits}, where the $\chi^2/{\rm d.o.f.}$ and $p$-values of the constant fits are also shown.  For some of these fits, especially the $p_2$ bound state fits, the $\chi^2/{\rm d.o.f.}$ values are larger than one would expect, suggesting that the systematic errors are underestimated.  There is no clear trend in the volume dependence of these taste-splittings.  Rather there is larger random scatter about a constant than would be expected by chance alone, so we inflate the fit error by a scale factor equal to $\sqrt{\chi^2/{\rm d.o.f.}}$ in those instances where this quantity exceeds one.  This is a rough way of accounting for underestimated systematic errors when averaging disparate results, and it is the procedure adopted by the Particle Data Group \cite{ParticleDataGroup:2024cfk} when the results of different experiments included in an average are in tension.

Note that the weak dependence of the taste-splittings on volume disfavors the interpretation where $M_\chi$ is set by the Hubble scale of the lattice geometries.  The Hubble scale on these lattices is not much smaller than the Planck scale, where it is either comparable to or smaller than by a factor of a few, so this serves as a non-trivial test of the dependence of the taste-splittings on the curvature scale of the semiclassical background.


\begin{center}
\begin{table}
\caption{Summary of results shown in Fig.~\ref{fig:Bandplots} for constant fits to various taste-splittings versus volume.  The first column is $\kappa_2$, which labels the lattice spacing, and the second is the $p_i$ block labeling the taste-splitting.  The third column is the result of the fit to the taste splitting, including its statistical error.  The fourth column is the $\chi^2/{\rm d.o.f.}$ of the fit, and the fifth column is its corresponding $p$-value.  All fits with $\chi^2/{\rm d.o.f.}>1$ have their errors rescaled by $\sqrt{\chi^2/{\rm d.o.f.}}$.}
\label{tab:band_fits}
\begin{tabular}{c c c c c} 
\hline \hline
$\kappa_2$& $p_{n}$ & $M_{\chi}$ & $\chi^{2}/{\rm d.o.f.}$ & $p$-value \\
\hline
\multirow{3}{4em} {\centering 2.45}& 
$p_{1}$ & 0.2868(73) & 0.922 & 0.450\\
& $p_{2}$ & 0.1848(24) & 0.467 & 0.760\\
& $p_{3}$ & 0.232(26) & 0.508 & 0.730\\ 

\hline
\multirow{3}{4em} {\centering 3.0}& 
$p_{1}$ & 0.151(13) & 0.440 & 0.821\\
& $p_{2}$ & 0.0678(56) & 2.234 & 0.048\\
& $p_{3}$ & 0.143(26) & 1.883 & 0.094\\
\hline
\multirow{3}{4em} {\centering 3.4}& 
$p_{1}$ & 0.0844(54) & 0.262 & 0.903\\
& $p_{2}$ & 0.0046(43) & 1.670 & 0.138\\
& $p_{3}$ & 0.067(9) & 0.843 & 0.519\\
\hline
\multirow{3}{4em} {\centering 3.8}& 
$p_{1}$ & 0.011(26) & 1.792 & 0.146\\
& $p_{2}$ & -0.0118(64) & 3.512 & 0.0036\\
& $p_{3}$ & 0.0424(28) & 0.244 & 0.943\\
\hline\hline
\end{tabular}
\end{table}
\end{center}


\subsection{Continuum extrapolation}

\begin{figure}[h]
    \centering
    \includegraphics[width=1.2\linewidth]{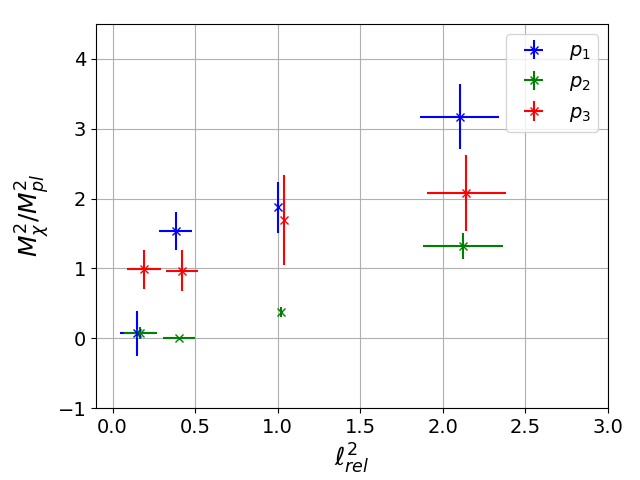}
    \caption{Taste splittings in the chiral limit expressed in Planck units versus $\ell_{\rm rel}^2$ for the $p_1$, $p_2$, and $p_3$ states.  Horizontal offsets are introduced for the $p_{2}$ and $p_{3}$ data sets.}
    \label{fig:continuum}
\end{figure}

We are interested in the behavior of the non-zero taste-splittings as we take the continuum limit.  The trend in $M_\chi$ as the continuum limit is approached can be seen more clearly in Fig.~\ref{fig:continuum}, where we plot $M_\chi^2$ in units of the Planck mass versus the (dual) lattice spacing squared for the three non-zero taste-splittings, analogous to what is usually done for pion taste-splittings in Lattice QCD with staggered fermions \cite{MILC:2009mpl}.  One can see that the $p_2$ splitting goes to small values, consistent with zero, at our smallest lattice spacings.  The $p_1$ and $p_3$ splittings also trend smaller, though they are noisy, making a definitive continuum extrapolation challenging.  Our tentative picture is that the taste-splittings decrease slowly with decreasing lattice spacing, and then more rapidly once the splittings drop below $\sim M_{\rm pl}$.  This is supported by the fact that the $p_2$ bound state is around $M_{\rm pl}$ on our coarsest lattice spacing and decreases faster than linear in $\ell^2$ as the continuum limit is approached.  The $p_1$ and $p_3$ splittings start well above the Planck scale and decrease more slowly, though the $p_1$ appears to cross this threshold at our finest lattice spacing, where its value drops significantly.  Runs at finer lattice spacings would be especially helpful in order to test this picture.

\begin{figure}[h]
    \centering
    \includegraphics[width=1.2\linewidth]{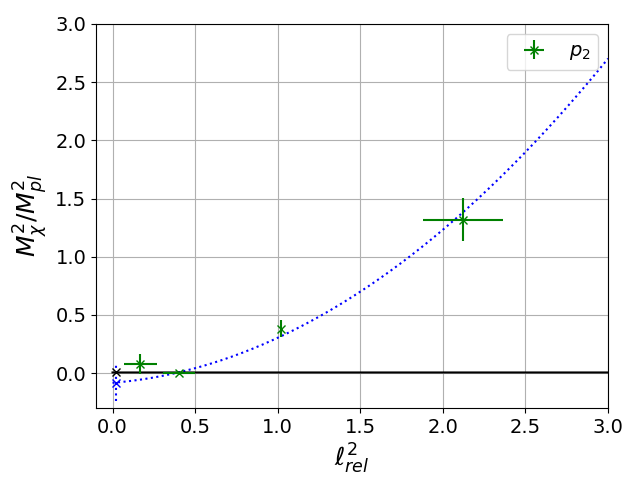}
    \caption{The $p_2$ taste splitting (in Planck units) versus $\ell_{\rm rel}^2$, with a constant fit (solid black) to the finest two lattice spacings and a quadratic fit (dotted blue) to all four data points.  Continuum extrapolated values for both fits with statistical errors are also shown.}
    \label{fig:continuum_p2}
\end{figure}

Although we are not yet able to resolve the question as to whether $M_\chi$ approaches a value ${\cal O}(M_{\rm pl})$ or vanishes in the continuum limit, we do have suggestive results in favor of the latter.  We focus on the $p_2$ taste splitting, shown in Fig.~\ref{fig:continuum_p2}, along with fits that extrapolate to the continuum limit.  The finest two points are compatible with zero.  A constant fit to these points yields 0.005(7) with $\chi^2/{\rm d.o.f.}$ of 0.238 and $p$ value 0.870.  We also include a quadratic fit to all four points, which extrapolates to slightly below zero (but consistent with zero within statistical errors) and has a $\chi^2/{\rm d.o.f.}$ of 3.22 and an acceptable $p$ value of 0.073.  Thus, the $p_2$ taste-splitting does not appear to be ${\cal O}(M_{\rm pl})$ in the continuum limit, given the small size to which it is constrained.  If one taste splitting vanishes in the continuum limit, it is likely that the others do as well, though the others are not as well constrained by our current simulations.  

\begin{figure}[h]
    \centering
    \includegraphics[width=1.2\linewidth]{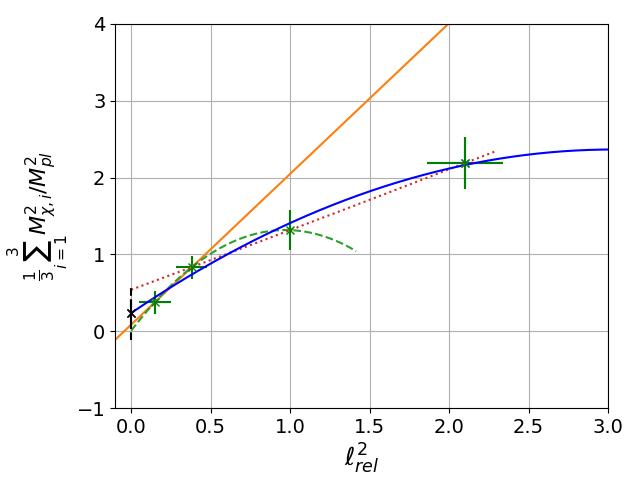}
    \caption{The average of the three non-zero taste splittings versus $\ell_{\rm rel}^2$. Also shown are various continuum extrapolation fits, including a quadratic fit through all four points (solid blue), quadratic fits dropping the first (dotted red) and last (dashed green) points, and a linear fit through the two finest lattice spacing points (solid orange).  The extrapolated value from the quadratic fit to all points is included with statistical error (solid riser line) and an estimate of the continuum extrapolation systematic error included in quadrature (dashed riser line).  Fits including only the finest lattice spacings are close to zero in the continuum limit.}
    \label{fig:continuum_avg}
\end{figure}

As another test of taste-symmetry restoration, we consider the average of the three non-zero splittings, since we expect an average to reduce random noise, making it easier to see a systematic trend.  Figure~\ref{fig:continuum_avg} shows $M_\chi^2$ (again in Planck units) averaged over the three tastes, versus $\ell_{\rm rel}^2$.  We extrapolate to the continuum using a quadratic fit, and we estimate an extrapolation systematic error from fit variations that omit either the first or last data points.  We also show a simple linear extrapolation to our finest two data points.  Given the large errors, it is difficult to say whether the average taste-breaking vanishes or extrapolates to some not so small faction of the Planck mass, since it is compatible with both.  We see from Fig.~\ref{fig:continuum_avg} that the fits including only the finest two or the finest three data points have continuum limits that are quite close to zero.  This suggests that taste symmetry is restored in the continuum limit, though results at even finer lattice spacings would be needed to make a stronger statement.  Even so, the current analysis is a significant improvement over the exploratory work of \cite{Catterall:2018dns}, where neither of our two finest lattice spacings was available.


\section{Conclusion}
\label{sec:conclude}

This work presents a study of the two-fermion bound state spectrum of K\"ahler-Dirac fermions coupled via gravity within the EDT formulation.  The $U(1)$ symmetry ensures that some of the taste bound states vanish in the chiral limit, and our numerical studies confirm this is the case.  The states that are not protected by that symmetry do not vanish in the chiral limit, leading to non-zero taste-splittings, the behavior of which is the focus of the present work.  This study improves over the exploratory study of \cite{Catterall:2018dns} by including two finer lattice spacings and several volumes at each lattice spacing, as well as the use of more reliable lattice spacing determinations on these ensembles \cite{Dai:2024vjc}.  We find that the average over non-vanishing taste-splittings, i.e. the squared masses $M_\chi^2$ of the various two-fermion bound states that do not vanish in the chiral limit, decreases as the continuum limit is taken.  Although the $p_1$ and $p_3$ bound states are noisy, the $p_2$ state is relatively clean, and there is solid evidence that it vanishes in the continuum limit.  This suggests that the taste-breaking scale is not set by the Planck scale.  Furthermore, the weak dependence of the taste-splittings on volume suggests that the taste-breaking scale is also not set by the curvature scale.  Rather, the taste-breaking appears to be a lattice artifact, with the taste symmetry being restored in the continuum limit without fine tuning.  This provides evidence that K\"ahler-Dirac fermions coupled gravitationally via EDT is in the same universality class as (four flavors of) Dirac fermions with unbroken chiral symmetry.  This is non-trivial, since K\"ahler-Dirac fermions in the presence of a curved classical background is not equivalent to four copies of Dirac fermions.

One feature that is new with K\"ahler-Dirac fermions coupled via EDT compared to flat-space lattice theories is the observation that the different blocks of the diagonal mass matrix appearing in the K\"ahler-Dirac action require rescaling factors for the bare masses to ensure that the various taste copies are degenerate in the continuum limit.  Note that no additive renormalization seems to be required, since the limit $m_0\to 0$ in each block is enough to recover the chiral limit, subject to the usual requirement that the continuum limit is taken first.  Thus, a multiplicative renormalization for the mass in each block appears to be sufficient to recover taste symmetry, if taste-symmetry is indeed restored in the continuum limit.  A further check of this is that the $p_0$-block mass rescaling required to match the $p_4$-block mass can be identified as the ratio of the direct and (renormalized) dual lattice spacings, which was previously determined from demanding that the EDT lattice geometries recover classical de Sitter space in the large volume limit.  That the taste bound states give the same value for this ratio is a good consistency check, providing further evidence that our simulations recover de Sitter space.

The current analysis could be improved with more statistics on the ensembles used in this work, as well as an effort to improve the determinations of the lattice spacings.  The generation of an even finer lattice spacing is in progress, and this would further improve the extrapolation to the continuum limit.  In addition, future studies that deal with fermions consistently will need to be unquenched, with the fermion back-reaction included in the generation of the geometries.  An effort to do this is also underway, and exploratory unquenched runs are in progress.  It is our future goal to repeat the present analysis unquenched, with the back-reaction of the fermions on geometry properly taken into account.


\begin{acknowledgments}
The authors thank Simon Catterall for valuable discussions.  WF, and JL were supported by the U.S. Department of Energy (DOE), Office of Science, Office of High Energy
Physics under Award Number DE-SC0009998.  MG was supported by a Syracuse University SOURCE Bridge Award.  MD was supported by the U.S. Department of Energy (DOE) and Office of Science under Award Number DE-SC0015655.
JUY was supported by the U.S. Department of Energy grant DE-SC0019139, and by Fermi
Research Alliance, LLC under Contract No. DE-AC02-
07CH11359 with the U.S. Department of Energy, Office
of Science, Office of High Energy Physics, and by the Department of Energy through the Fermilab Theory QuantiSED program in the area of ``Intersections of QIS and Theoretical Particle Physics.''
The research of MS was supported by an NWO Veni grant under grant ID  [\url{https://doi.org/10.61686/SUPEH07195}].  Computations were performed in part on the Syracuse University HTC Campus Grid and were supported by NSF award ACI-1341006. This research was enabled in part by support provided by ACENET, Calcul Québec, Compute Ontario, and the Digital Research Alliance of Canada. The authors acknowledge support by the state of Baden-Württemberg through bwHPC.
\end{acknowledgments}


\bibliography{apssamp}

\end{document}